\documentstyle[prl,aps,epsf,psfig]{revtex}
\newcommand \beq{\begin{eqnarray}}
\newcommand \eeq{\end{eqnarray}}
\newcommand \be{\begin{eqnarray}}
\newcommand \ee{\end{eqnarray}}
\newcommand{\set}[2]{\newcommand{#1}{#2}}
\set{\pa}{\partial \over \partial\, }
\set{\leftvector}{\stackrel{\leftarrow}{\partial }}
\set{\rightvector}{\stackrel{\rightarrow}{\partial }}
\begin{document}
\twocolumn[\hsize\textwidth\columnwidth\hsize
           \csname @twocolumnfalse\endcsname
\title{Electric field dependence of pairing temperature and tunneling}
\author{K. Morawetz}
\address{Max-Planck-Institute for the Physics of Complex Systems, 
Noethnitzer Str. 38, 01187 Dresden, Germany
}
\date{\today}
\maketitle
\begin{abstract}
Using the Bethe-Salpeter equation including
high electric fields, the dependence of the critical temperature of
onsetting superconductivity on the
applied field is calculated analytically. The critical
temperature of pairing is shown to increase with the applied field strength. 
This is a new field effect and 
could contribute to the explanation of recent experiments on field induced
superconductivity. From the field dependence of
the Bethe-Salpeter equation, the two--particle 
bound state solution is obtained as a
resonance with a tunneling probability analogous to the 
WKB solution of a single particle confined in a potential and coupled
to the electrical field. 
\end{abstract}
\vskip2pc]

The influence of static electric fields on superconducting properties
has been investigated for nearly 40 years \cite{GS60} but
has gained renewed experimental interest
\cite{MBCAHM94,SOKLCCG97,WHKG99}, see also overview in \cite{KS98}.
Recently, field-induced
superconductivity in a spin-ladder cuprate, where the critical
temperature raised to 14 K by applying high gate voltages has been reported \cite{S01}.
A wide experimental activity on high-$T_c$ superconductors
has been devoted to this change of the critical temperature and transport properties due to an external electric field \cite{Ma92,MSBG93,MBCAHM94,KSKTB97,SOKLCCG97,ABHS98}.
Consequently a considerable theoretical effort has been made to describe such field effects \cite{Sa93,BKS93,UGD02}.

Usually two standard mechanisms have been
proposed in high $T_c$-cuprates \cite{FMBW95}, see \cite{KS98} and citations
therein. The first one describes the changes of hole concentration due
to Coulomb force of the external field and the second mechanism
describes the field-induced oxygen rearrangement \cite{CVG94}. 
These mechanisms
will change the density of state and consequently the order parameter.
As a result one expects frustration of charge density waves
\cite{HY00} connected with specific threshold electric fields
\cite{DVM01} and a nonlinear conductivity \cite{MPI01}. However, the
nonlinear field dependence of the current showed that there must be a
field effect besides the change of carrier density \cite{WHKG99}.

For very clean and thin two-dimensional structures with high
transversal electric fields one can expect that the pairing mechanism
by itself and the two-particle correlations will be influenced by the
field since the direct current as the dominant effect of the field
is suppressed by the geometry. This was first investigated
in a quasi-one-dimensional conductor \cite{gM96}. For general
dimensions one best work within the field dependent Bethe-Salpeter equation.
We will find that the pairing mechanism is affected itself analogously to the
formation of sidebands in the density of states due to high fields for one
particle properties \cite{BJ92}. Here we will show that the
two-particle Bethe-Salpeter equation becomes field dependent and
consequently also the
resulting critical temperature.
In an earlier paper \cite{MOR94} it was demonstrated that the inclusion of high electric
fields leads to novel modifications of the two-particle properties
: (i) The bound
states, reflected by the value $\pi$ of the phase shift at low
energies are turned into resonances, (ii) The Pauli-blocking
effect (which turns the phase shift to negative values,
indicating effective repulsive behavior of the scattering) is
suppressed by the applied field. This is understood as the
opening of a scattering channel due to the field. (iii) The onset
of pairing and superfluidity (which is reflected as a sharp resonance at
energies twice the chemical potential) is observed at
higher temperatures with increasing field. 

Here in this letter we will concentrate on the latter effect and will derive an analytical expression of the increase of the critical temperature with the applied field. In order to see the expected field effects we consider the two-particle phase factor
for free charged particles in a constant electric field ${\bf E}$ [$t=(t_1+t_2)$, $\tau=t_1-t_2$]
 \begin{eqnarray*}
&&{\rm exp}
\left (
- {i \over \hbar}
\int\limits_{t_1}^{t_2} dt
\left (
\frac{({\bf p} + e_1 {\bf E} t )^2}{2 m_1}+
\frac{({\bf-p} + e_2 {\bf E} t )^2}{2 m_2}
\right )
\right )
\nonumber\\
&&={\rm exp}
\left [
{i\over \hbar} \tau
\left (
\frac{p^2}{2 \mu_{12}} - \frac{E^2}{2}
\left (
\frac{e_1^2}{m_1} + \frac{e_2^2}{m_2}
\right )
t^2
\right )
\right .
\nonumber\\
&&\left .
-{i \over \hbar}
\left (
\frac {e_1}{ m_1} - \frac{e_2}{ m_2}
\right )
{\bf p E } t \tau -i \frac {E^2}{ 8\hbar}
\left (
\frac {e_1^2}{ m_1 }+\frac {e_2^2}{ m_2}
\right )
\frac {\tau^3}{ 3 }
\right ],
\end{eqnarray*}
where three different actions of the applied field occur. The first
term proportional to $t^2$ can be absorbed into a redefinition of
the chemical potential, i.e.\ the background. The second term,
proportional to $t$ and linear in the field, is vanishing for equal
mass to charge ratios as considered here. The last term,
proportional to $E^2$, is linked to the third power of the
difference time $\tau$ and is therefore due to off-shellness. It is not vanishing for equal charge--to--mass ratios and
will turn out as the essential term responsible for the here described
modification of pairing and tunneling. The typical energy scale which appears here is
\beq\label{nu}
{ \lambda_E=\left( \frac{E^2\hbar ^2}8(\frac{e_a^2}{m_a}+\frac{e_b^2%
}{m_b})\right) ^{\frac 13}}.
\label{le}
\eeq

In the following we
proceed to investigate the two-particle properties more closely by
solving the Bethe-Salpeter equation. We intend to show that the
applied electric field is leading to a new field effect to the
critical temperature. Therefore we do not explain how
superconductivity is occurring or which inhomogeneity of the gap might
be necessary \cite{Sca95,TK00} but concentrate on the change of the
critical temperature provided we do have superconductivity. 

In \cite{MOR94} we have shown that the gauge--invariant formulation
of the one-- and two--particle Green's function leads to a field,
${\bf E}$,
and time, $t$, dependent Bethe--Salpeter equation for the ${\cal
  T}$--matrix
with the difference momenta of incoming channel ${\bf p}$ and of
outgoing channel ${\bf p'}$,
which can be solved with separable
potentials
$
V(p,p^{\prime })=\lambda g(p)g(p^{\prime })
$. For our case of equal particles it reads
as
\beq
&&\langle {\bf p}|{ {\cal T}}^R({\bf K},\omega ,t)|{\bf
p^{\prime }}%
\rangle 
=\frac{\lambda g({\bf p^{\prime }})g({\bf p})}{1\!-\!\lambda J({\bf K},{\bf E},\omega ,t)},
\label{solution}
\eeq
with
\beq\label{solution1}
&&J
=\int \frac{d{\bar {{\bf p}}}}{(2\pi \hbar
  )^3}g({\bar {{\bf p}}})^2 
{\Phi (\varepsilon_a+\varepsilon_b\!-\!\omega )}\left( 1\!-\!f_{\frac{{\bf K}}2\!-\!{\bar {{\bf p}}}
}^a\!-\!f_{\frac{{\bf K}}2\!+\!{\bar {{\bf p}}}}^b\right)
\eeq
where the arguments of the quasiparticle energy $\varepsilon_{a/b}$
are the same as the one of
the
distribution functions $f^{a/b}$.
The center of mass momentum is ${\bf K}$ and
the function $\Phi $ is given in terms of Airy-functions,
\cite{a84}
\beq\label{phi}
{ \Phi (\omega)\!=\!\frac \pi { \lambda _E}\left( {\rm Gi}(\frac \omega 
{ \lambda _E})\!+\!i{\rm Ai}
(\frac \omega { \lambda _E})\right)} \rightarrow {1 \over \omega
\!+\!i 0} \, {\rm for} \, { E}\!=\!0.   \label{zitat}
\eeq
The usual Bethe-Salpeter equation is recovered in the field--free case.
The occupation factors $f^{a/b}$ are the 
one-particle distribution functions resembling the
Pauli-blocking of the intermediate propagator. At equilibrium we
have correspondingly Fermi distributions.

The denominator of the ${\cal T}$-matrix determines the resonances or
bound states which may occur in the system. 
Due to the equal charge--to--mass ratios, in which case the center
of mass time dependence drops out of the equation, the only remaining
center of mass time dependence is in the center of mass momentum
${\bf K}$ but which is suppressed due to geometric quasi-2D restriction.

In order to demonstrate the field effects, we use the model of a contact
potential. This
is most easily obtained from the separable potential 
by the limit of zero range, which means 
$g(p)\to\infty$, in such a way that the scattering length is
reproduced. This is accomplished by adapting $\lambda$ accordingly. 
Therefore  the scattering amplitude is normalized properly
to render finite results. The definition of the scattering length $a$ 
is given by the small wavelength expansion of the scattering phase shift 
\beq
p {\rm cot} \delta &=& p {{\rm Re}{\cal T}\over {\rm Im}{\cal T}}=p
{1-{\lambda} {\rm Re} J_0(\varepsilon(p))\over {\lambda} {\rm Im}
  J_0(\varepsilon(p))}
\approx -{\hbar \over a}
\label{phi1}
\eeq
which we use to replace the strength $\lambda$ of the potential by the scattering length
and perform the limit to the contact potential by $g(p) \to \infty$. The
${\cal T}$-matrix reads then
\beq
{\cal T}_{pp'}(\omega)&=&{g(p) g(p') \over {\rm Re} J_0(\varepsilon(p))-{\hbar
    \over a p} {\rm Im} J_0(\varepsilon(p)) -J(\omega)}\nonumber \\
&{\rightarrow}& {1\over {\rm Re} J_0(\varepsilon(p))-{\hbar \over a p}
  {\rm Im} J_0(\varepsilon(p)) -J(\omega)}\label{tm},
\eeq
with $J(\omega)$ given by (\ref{solution1}) but without form factor $g(p)\to
1$
and $J_0=J[f\to 0]$.
Now we search for a pairing resonance which appears as a jump in the
phase shift by $\pi$ at twice the chemical potential, $2 \mu$, in the
field free case. For the field dependent case there will be a shift of
this two particle threshold $2 \mu+\zeta$ according to the Stark
effect. The conditions for such a pairing resonance  is that $\tan\delta$
has a pole with vanishing strength. From (\ref{phi1}) and 
(\ref{tm})
follows
\be
{\rm Re} J_0(\varepsilon(p))-{\rm Re}J(2 \mu+\zeta)&=&{\hbar \over a p} {\rm Im
  }J_0(\varepsilon(p))\nonumber\\
{\rm Im}J(2 \mu+\zeta)&=&0.
\label{sys}
\ee
In the field free case we have from (\ref{tm}) ${\rm Im} J_0=N(p) \pi$
with the density of states $N(p)$. The real part diverges by
itself, but the difference ${\rm Re}(J-J_0)$ remains finite. This can
be either realized by a finite range of potential and a corresponding
finite form factor $g(p)$, or, alternatively, by an
energy cutoff $\omega_c\ll \mu$. We will adopt here the later possibility  since it
compares to the standard BCS treatment. Since ${\rm Im}J_0(2\mu)=0$, we
obtain from (\ref{sys}) with $V_0=4 \pi a \hbar^2/m$ the usual BCS equation for the critical
temperature
\be
N(p_f)\int\limits_0^{\omega_c} {d\xi\over 2 \xi} \tanh{\xi\over 2 T_0}=-{1\over V_0}.
\ee

For the field dependent case we have to solve the coupled equation
system (\ref{sys}) for the energy shift $\zeta$ and the critical field
$T_0$. Further we allow a nontrivial realistic density of states $N$, for instance due to layered
structures. According to (\ref{solution1}) and (\ref{zitat}) the kernel of ${\rm Im} J$ is the
Airy function. Expanding the second equation of (\ref{sys}) in terms of the
field parameter $\lambda_E$ we obtain to the lowest order,
$\zeta=-{\lambda_E^3/ 24 T_c^2}+o(\lambda_E)^6,
$
analogously to the quadratic Stark effect.
We will now solve the equation for the critical temperature and will
find that it leads to results $\sim\lambda_E$. Therefore we can neglect
the influence of $\zeta$ on the first equation of (\ref{sys}), it being
of higher order, and
obtain with (\ref{zitat})
\be
&&-{1\over N(p_f) V_0}=\int\limits_{-\omega_c}^{\omega_c}d\xi {\pi\over
  \lambda_E} {\rm Gi}({2\xi\over
  \lambda_E}) \tanh{{\xi\over 2 T_c}}=
\nonumber\\
&&\int\limits_{0}^{\omega_c}d\xi  
\left [
{1\over 2 \xi}\!-\!
{\pi^{1/2} \over { (2\xi \lambda_E^3)^{1/4}}
}
 \cos{
\left (
\frac 2 3
      \!\left (
{2 \xi\over \lambda_E}
      \right )^{3/2}
\!\!+\!\!{\pi\over 4}
\right )
}
\right ]
\tanh{\xi\over 2 T_c},
\nonumber\\
&&
\label{tce1}
\ee
where we have used the asymptotic expansion for the $Gi$ function for
small $\lambda_E$. We remark that they are different above and below
the Fermi energy in contrast to the field-free case. Please not also
that this field dependent modification can be considered as an
effective field dependent density of state. 
This shows that the Anderson theorem
\cite{Ric69a} is fulfilled which states, that for a homogeneous
perturbation and order parameter the critical temperature can only be
effected by the density of states. 

Since
$\omega_c/T$ is very large, the first part of (\ref{tce1}) can be
integrated in the standard manner \cite{Ric69},
\be
&&\int\limits_0^{\omega_c} {d\xi\over
  2 \xi}\tanh{{\xi\over 2 T_c}}=\frac 1 2 \ln{{\omega_c\over T_c}} [1-2 f(\omega_c)]\nonumber\\
&&+
\int\limits_0^{\omega_c/T_c} d x \ln{x}{d \over d x} \left ({1\over
    {\rm e }^x+1}\right )
\approx \frac 1 2 \ln{{a \omega_c\over T_c}}
\label{tc2a}
\ee
and $a=2 {\rm e}^\gamma /\pi\approx 1.134$.
The second part of (\ref{tce1}) can be expanded up to $o(\lambda_E/T_c)^5$
\be
&&{\lambda_E\over 8 T} \sqrt{\pi}\! \int\limits_0^{2\omega_c/\lambda_E}\!\! d p
p^{3/4} \cos{\left (\frac 2 3 p^{3/2}\!+\!{\pi\over 4}\right )}
=-{\lambda_E\over T} b
\ee
with $b={3^{2/3} \sqrt{\pi}\over 2^{1/6}}
 \left ({1\over 32}\Gamma\left (\frac 7 6 \right )+{3\over 112}
\Gamma\left ({13\over  6}\right )\right )\approx0.190$.
Using the result for the field--free case $T_0[N]=a \omega_c \exp{(-2/|V_0|
  N)}$ dependent on the doping concentration, we obtain finally the
field--dependent critical temperature
\be
{T_c} \ln{T_c\over T_0[N]}=b\lambda_E+o(\lambda_E/T_c)^5
\label{sol}
\ee
from which one gets $T_c-T_0[N]\approx \lambda_E b$ for small
fields. Please remark that the doping concentration condensed in 
the density of states $N$ is field dependent
itself such that Eq.(\ref{sol}) is an effect on top of the doping
dependence which might be an explanation for the nonlinear dependence
found in \cite{MBCAHM94,WHKG99}.
Within the regime
$\lambda_E\ll T_c\ll\omega_c\ll \epsilon_f$ we find an increase of
$T_c$ with the applied field. For other expansions of (\ref{tce1}) one
finds a decrease, like when $\lambda_E\sim \omega_c$ which has been
reported in \cite{gM96}.
The result here, (\ref{sol}), represents a new field effect. 
All field effects considered so
far in the literature due to the increase of
doping or the change of bands are condensed in the above $T_0[N(p_f)]$ parameter.  

\begin{figure}
\psfig{file=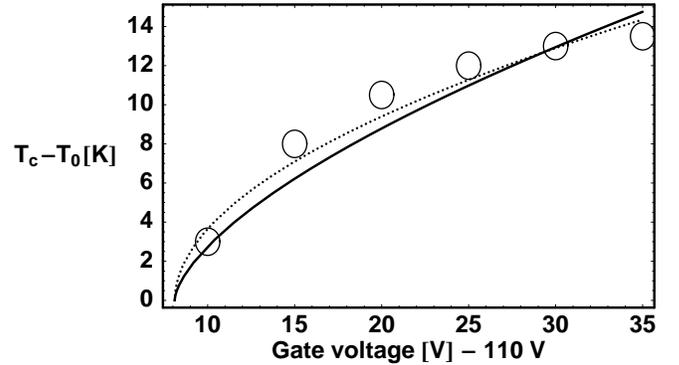,width=9cm}
\caption{ The critical temperature in dependence on the
  applied voltage according to (\protect\ref{sol}). The circles gives
  the experimental values of \protect\cite{S01}. The solid line gives
  the theoretical value of (\protect\ref{sol}) for $\epsilon_{\rm
  ins}/\epsilon_{{\rm cup}}\approx 10$ and $T_0=2\times 10^{-8}$ K
while the dashed line is for $\epsilon_{\rm
  ins}/\epsilon_{{\rm cup}}\approx 192$ and $T_0=1$ K. }
\label{tce}
\end{figure}

In figure~\ref{tce} we compare the solution of (\ref{sol}) with the
experimental values of \cite{S01}. This experiment has been performed 
on a field effect transistor device of
[CaCu$_2$O$_3$]$_4$ films  on a MgO substrate covered by an insulating
Al$_2$O$_3$ sheet. The capacitance
of this insulating sheet is 120 nF/cm$^2$. With a 
ratio of dielectric constants between the insulator and the cuprate of
$\epsilon_{\rm ins}/\epsilon_{{\rm cup}}$, the effective electric field
strength in the active sheet depends on the applied voltage $U$ as 
$0.135 \epsilon_{\rm ins}/\epsilon_{{\rm cup}} \times U/$nm. 
We assume that above
a critical voltage sufficient charges are doped into the cuprate surface
layer such that correlations and pairing occurs. This threshold
voltage was at about 118V assuming $T_0=1K$. Above this threshold the
field effects act as
derived above. The best fit of the dielectric constants between insulator and
cuprates are found here $\epsilon_{\rm ins}/\epsilon_{{\rm cup}}\approx 192$. 
Nearly the same figure appears if we set  $\epsilon_{\rm
  ins}/\epsilon_{{\rm cup}}\approx 10$ and $T_0=2\times 10^{-8}$ K
which is probably more realistic since $\epsilon_{\rm ins}/\epsilon_{{\rm cup}}=0.7$ \cite{pw}.

The field dependence changes directly
the two--particle correlations, which leads to an increase of the critical
temperature $\sim U^{2/3}$ of the applied voltage. Of course, at higher
voltages the breakthrough occurs limiting this effect. A more
realistic calculation would be to solve (\ref{tce1}) directly 
as well as
realistic density of states $N(p_f)$. This will lead to a nonlinear
decrease of the curve in figure~\ref{tce} for higher applied fields
and is devoted to a forthcoming work.

Here it has to be remarked that the discussed experimental values do
not allow to conclude uniquely on this new field effect. As long as
the doping concentration $N$ is not simultaneously measured one cannot
decide how much field effect is on top of the conventional $T_0[N]$
result according to (\ref{sol}).

In order to gain some trust into the theoretical treatment above we
want to investigate now possible bound states of the field
dependent Bethe Salpeter equation. A bound state, parameterized
by a negative (attractive) scattering length, should turn into
a resonance with finite lifetime when the field is applied.

Neglecting the medium effects condensed in the occupation
numbers, we obtain with quadratic dispersion from (\ref{solution1}) 
after some analytical work for 3D
\beq
{\rm Im} \,J(z)&=&{m \sqrt{{ m \lambda_E}} \over  \hbar^3 2^{5/3}}
({\rm Ai}'^2(z)-z {\rm Ai}^2(z))\nonumber\\
{\rm Re} \,J(z)&=&{m \sqrt{{ m \lambda_E}} \over  \hbar^3 2^{5/3}}
({\rm Ai}'(z){\rm Bi}'(z)-z {\rm Ai}(z){\rm Bi}(z)),
\label{imre}
\eeq
with
$z=- {p^2 \over m 2^{2/3} { \lambda_E}}$ for
$J({p^2\over m})$ and
$z={|\omega| \over
2^{2/3} { \lambda_E}}$ for $J(\omega)$, respectively,
and standard Airy functions
.
Together with (\ref{tm}), this represents the result for the field
dependent Bethe-Salpeter equation in high electric fields. Analogous
formulae can be given for 2D. The negative poles of this equation provides us with the influence of the field on the bound states. The latter
one is realized by choosing the attractive contact potential in a way
that the negative squared scattering length (\ref{phi1}) 
reproduces the bound state
energy $\omega=-E_0=-\hbar^2/m a^2$.
It is interesting to see that for
vanishing fields, $\lambda_E \rightarrow 0$, we get  from (\ref{imre})
that ${\rm Im }J({p^2\over m})\to {m p/
4 \pi \hbar^3}$ and ${\rm Re }J(\omega)\to
{-m^{3/2} \sqrt{| \omega|} /4 \pi \hbar^3}$ while ${\rm Re }J({p^2\over m})= {\rm Im }J(\omega)\to 0$,
which leads from (\ref{tm}) to the known field--free on-shell ${\cal T}$-matrix
${\cal T}_{E=0}({p^2\over m})=-{4 \pi \hbar^3 \over m}/ \left ({\hbar \over a}+i p\right )
$.

For small fields we investigate now the bound states and expand
(\ref{tm}) up to $o(\lambda_E^3)$ field effects.
The bound or resonance states are given by the pole of the ${\cal
  T}$-matrix and we must search for the complex zeros of $\omega$ via
\beq
&&{\cal T}^{-1} (\omega){4 \pi \hbar^3 \over m^{3/2}}=-{\hbar\over
  a\sqrt{m}}+\sqrt{|\omega|}-i {\lambda_E^{3/2} \over 4 |\omega|} 
{\rm e}^{-\frac 2 3 \left ({|\omega| \over \lambda_E}\right)^ {3/2} }
=0.
\label{eb0}
\nonumber\\
&&\eeq
The
pole-value of the T-matrix is now a resonance
$
|\omega|=-E_0+\Delta +i {\hbar \over \tau},
\label{eb}
$
and reads in the lowest expansion $o(\lambda_E)^{7/2}$
\beq
\Delta&=&{{ \lambda_E}^3 \over 16 E_0^2}\left (3+4 \left ({-E_0\over
      \lambda_E}\right )^{3/2} \right )
\exp{\left (-\frac 4 3 \left ({-E_0 \over { \lambda_E}}\right )^{3/2}\right )}\nonumber\\
{\hbar \over \tau}&=&{{ \lambda_E}^{3/2} \over 2 \sqrt{-E_0}} \exp{\left (-\frac 2 3 \left ({-E_0 \over {\lambda_E}}\right )^{3/2}\right )}
\label{eb1}.
\eeq
If one particle is very heavy, $m_b\to \infty$, and counting the bound
state energy $-E_0=U_0-\epsilon$ from the continuum threshold $U_0$, we
obtain from (\ref{eb1})
\beq
{\hbar \over \tau}\sim \exp{\left (- \frac 4 3 {\sqrt{2m} \over e\hbar E} (U_0-\epsilon)^{3/2}
\right)}
\eeq
which is exactly the known WKB result 
for one particle
tunneling through a potential wall modified by an external field
$U_0-E x$.
Consequently, 
(\ref{eb1}) generalizes the known
WKB result towards the  two-particle problem in electric fields.
The bound state
energy and the damping or
inverse lifetime (\ref{eb1}) increases with increasing fields.
%
%

To summarize, we have analyzed the two--particle Bethe-Salpeter equation
with respect to the critical temperature of pairing and bound states
when an external electric field is applied. The critical temperature
rises with the applied field strength. This establishes an isotropic field effect
directly on the pairing mechanism beyond the ones considered so far. 
The bound states turn into resonances where the life time and energy
shift of the two--particle bound state is given explicitly.
As another application we may also think of the pair creation in high fields as
it appears in the neighborhood of big charged nuclei as electron
positron production or as meson production when strings are
breaking. 

The discussions with Marco Ameduri, Peter Fulde, Richard Klemm and Kazumi Maki
 are gratefully acknowledged.

Note added in proof:\\
Recently the Lucent report by Bell Labs (http://www.lucent.com/news$_{-}$events/researchreview.html) found scientific misconduct in a series of experimental results. The data which are used in this paper here are not among the objected ones in this report and are a result of a broader cooperation. 
 

\end{document}